\documentclass[10pt, conference, letterpaper]{IEEEtran}
\usepackage{adjustbox}
\usepackage{enumerate}
\usepackage{graphicx}
\usepackage{epstopdf}
\usepackage{epsfig}
\usepackage{color}
\usepackage{url}
\usepackage{etoolbox}
\usepackage{booktabs}
\usepackage{amsmath}
\usepackage{amssymb}
\usepackage{mathtools}
\usepackage{hhline}
\usepackage{setspace}
\usepackage{balance}
\usepackage{xcolor,colortbl}
\usepackage{array}
\usepackage[ruled, noend]{algorithm2e}
\usepackage{blindtext}
\usepackage{multirow}
\usepackage{mathrsfs}
\usepackage{tabularx}
\usepackage{cleveref}
\usepackage{spverbatim}
\usepackage{float}
\usepackage{caption}
\usepackage{tabularx}
\usepackage{cite}
\usepackage{subcaption}
\usepackage{cleveref}

\crefname{section}{\S}{\S}

\definecolor{LightCyan}{rgb}{0.88,1,1}
\definecolor{Gray}{gray}{0.9}
\definecolor{ashgrey}{rgb}{0.7, 0.75, 0.71}

\usepackage[utf8]{inputenc}
 
 \IEEEoverridecommandlockouts

\IEEEpubid{\makebox[\columnwidth]{978-3-903176-27-0~\copyright{} 2020 IFIP \hfill} \hspace{\columnsep}\makebox[\columnwidth]{ }}
\begin{document}

\author{
	\IEEEauthorblockN{
		Konstantinos Solomos\IEEEauthorrefmark{1},
		Panagiotis Ilia\IEEEauthorrefmark{1},
		Sotiris Ioannidis\IEEEauthorrefmark{2},
		Nicolas Kourtellis\IEEEauthorrefmark{3}}
	\IEEEauthorblockA{
		\IEEEauthorrefmark{1}University of Illinois at Chicago, USA; 
		\IEEEauthorrefmark{2}FORTH,Greece; 
		\IEEEauthorrefmark{3}Telefonica Research, Spain\\
		\IEEEauthorrefmark{1}\{ksolom6,pilia\}@uic.edu;
		\IEEEauthorrefmark{2}sotiris@ics.forth.gr;
		\IEEEauthorrefmark{3}nicolas.kourtellis@telefonica.com
		}
}

\title{Clash of the Trackers: Measuring the Evolution of the Online Tracking Ecosystem}	
\maketitle
	\begin{abstract}

Websites are constantly adapting the methods used, and intensity with which 
they track online visitors. However, the wide-range enforcements of regulations such us GDPR and 
e-Privacy has forced websites serving EU-based online visitors to eliminate, or at 
least reduce, such tracking activity, given they receive proper user consent.
Thus, it is important to analyze the aftermath of such policies, record the evolution of this 
tracking activity, and assess the overall ``privacy health'' of the Web ecosystem.
This work makes a significant step towards this direction.
In this paper, we analyze the ecosystem of 3rd-parties embedded in top websites, which amass the majority of online tracking, through six time snapshots taken every few months apart, in the duration of the last few years.
We perform this analysis in three ways:
1) by looking into the network activity that 3rd-parties impose on each publisher hosting them,
2) by constructing a bipartite graph of ``publisher-to-tracker'', connecting 3rd-parties with their publishers,
3) by constructing a ``tracker-to-tracker'' graph connecting 3rd-parties who are commonly found in publishers.
We record significant changes through time in number of trackers, traffic induced in publishers, embeddedness of trackers in publishers, popularity and mixture of trackers across publishers.
In the last level of analysis, we dig deeper and look into the interconnectivity of trackers, and how this relates to potential cookie synchronization activity.

\end{abstract}

	\section{Introduction}
\label{sec:introduction}

Online users' privacy is constantly violated by leaks of their PII to unauthorized parties.
Users lose their anonymity due to web tracking via cookies~\cite{roesner2012,englehardt_ccs}, device or browser fingerprinting~\cite{eckersley2010, acar2013, nikiforakis2013} and cookie synchronization~\cite{olejnik2014, bashir2018diffusion, acar2014, papadopoulos2018www-usercost, papadopoulos2019cookie}.
In the last few years, different legislations and directives such as the e-Privacy~\cite{eprivacy}, the General Data Protection Regulation (GDPR)~\cite{GDPR} in EU, and the California Consumer Privacy Act (CCPA)~\cite{ccpa} in USA, were introduced in an effort to increase transparency in user tracking and help users with personal data management and privacy protection.
In particular, GDPR forces websites to stop such activity or receive informed consent from their online visitors for any potential tracking, data collection and processing they may do, and also for any data sharing they may do with 3rd-parties.

Recent studies~\cite{iordanou2018tracing, degeling2018we, dabrowski2019measuring, sorenser19,utz2019informed} have investigated the aftermath of GDPR and its effects on the online tracking ecosystem, and how websites may have reduced their tracking activity.
However, the web tracking ecosystem is constantly evolving and adapting to new blocking methods.
Trackers continue their aggressive activity across multiple domains, at times unbalanced depending on the type of website~\cite{agarwal2020differentialtracking-hyperpartisan}, and even employ cookie-less, machine learning-based methods to track users across their different devices~\cite{zimmeck2017, brookman2017, solomos2019ctd-raid}, all in the name of ``more effective ad-campaigns''.

In this work, we build on previous studies and methods on web tracking, and perform a first of its kind longitudinal study to measure this ecosystem's changes over the last few years.
We perform this analysis in three levels through time, using six crawls of top Alexa websites in time snapshots of a few months apart.
First, we look into the network activity that 3rd-parties impose on each publisher (1st-party) hosting them.
With this first-level analysis, we confirm existing reports that claim reduction in tracking, by measuring generic HTTP network activity from 3rd-parties.
Second, we construct ``publisher-to-tracker'' ($PT$) bipartite graphs, connecting 3rd-parties with their publishers.
With this second level of analysis, we employ graph mining tools and metrics such as clustering coefficient, density, degree centrality, coreness, etc., to study the graph properties of the six bipartite graphs.
We find that the structure of the tracking ecosystem with respect to embeddedness in publishers has not changed significantly through time,
and that top degree centrality trackers such as Google's suite (google-analytics, doubleclick, etc.), Facebook, AppNexus, Criteo, etc., dominate the ecosystem in all time snapshots, without losing their market share of publishers.
Moreover, we also identify top betweenness trackers such as Twitter and Adobe, which are not in the typical top degree list but have embedded themselves in central positions in the web ecosystem.

Third, we construct a ``tracker-to-tracker'' graph ($TT$), connecting 3rd-parties that are commonly found in publishers.
With this analysis, we construct $TT$ pairs from the $PT$ graphs that can reveal potential collaborations between 3rd-parties.
We compare these pairs with ground truth data from confirmed data sharing flows of cookie syncing ($CS$) pairs, in two different $CS$ datasets from past studies.
Proper cookie syncing flows between trackers are not easy to get as they require activity from real users or persona-based automated browsing to trigger the CS mechanism.
Interestingly, we identify a high overlap between $CS$ and $TT$ pairs ($\sim$47\%-81\% when compared to previous ground truth $CS$ datasets).
We propose that such information flows can be inferred from the $TT$ graphs with reduced cost in deployment and measurements, as they require only web crawling of 1st-parties.

	\section{Background and Related Work}\label{sec:rel_work}

\subsection{Web Tracking \& Graph Modeling}
Many works have focused on analyzing the web tracking ecosystem, its internal mechanisms and their impact on user's privacy. 
One of the first studies on web tracking, by Mayer and Mitchell~\cite{mayer2012}, investigated which information is collected by 3rd-parties and how users can be identified. 
Roesner et al.~\cite{roesner2012} studied the various tracking behaviors and measured the prevalence of trackers while Falahrastegar et al.~\cite{pam} measured the existence of cookie synchronization trackers.
Papadopoulos et al.~\cite{papadopoulos2019cookie} used a heuristic-based mechanism to detect information exchanged between advertisers and found that 97\% of the users are exposed to cookie syncing at least once, and that ad-related entities participate in more than 75\% of the overall cookie synchronization.

A plethora of studies investigate stateful tracking techniques (e.g.,~\cite{roesner2012, olejnik2014, englehardt_ccs, trackers}), and stateless techniques such as browser fingerprinting~\cite{eckersley2010, acar2013, acar2014, nikiforakis2013, nikiforakis_www15, panchenko2016}. 
Acar et al.~\cite{acar2014} investigated the prevalence of ``evercookies'' and the effects of cookie respawning in combination with cookie syncing. 
Englehardt and Narayanan~\cite{englehardt_ccs} conducted a large scale measurement study to quantify both stateful and stateless tracking in the web, and Lerner et al.~\cite{lerner2016} conducted a longitudinal study of 3rd-party behaviors and found that tracking has increased in prevalence and complexity over time. 

In general, by studying the graph network properties, one is able to understand the characteristics of the tracking entities, and dissect the ecosystem and its inner mechanisms.
In that direction, Kalavri et al.~\cite{kalavri2016like} built a 2-mode bipartite graph  based on real user traffic logs, and focused their analysis on the communities formed by the graph vertices.  
Their analysis showed that trackers are well connected to each other, since 94\% of them are in the largest connected component.
Urban et al.~\cite{urban2018unwanted} collected behavioral data 
from emulated users located in 20 EU countries and created a 
cookie synchronization graph that connects 3rd-parties that share information.
They reported that the number of trackers and the number of direct syncing 
connections decreased through time, since fewer 3rd-parties are present in 
the publisher domains (40\% less syncing connections). Also, based on the 
properties of their graph, they found that the structure of the ecosystem 
did not change significantly.
Similarly, Bashir et al.~\cite{bashir2018diffusion} constructed a 
cookie syncing graph and associated the graph metrics  with the existence of different tracking domains.

\subsection{GDPR Enforcement and Web Privacy}

The GDPR~\cite{GDPR} is a regulatory initiative by the EU to harmonize data protection laws between its member states, and specifies under which circumstances personal data may be processed. 
Since the regulation directly affects the web and the online advertising and tracking ecosystem, recent works have focused on investigating the state of the ecosystem, the evolution of the privacy policies and their impact on user's privacy from different perspectives~\cite{dabrowski2019measuring,utz2019informed,sanchez2019can}.
Iordanou et al.~\cite{iordanou2018tracing} collected data from users across 
EU and identified the directions of tracking flows inside EU.
They reported that 85\% of the tracking flows terminate in servers located inside the EU, and identified that the most sensitive types of user information based on GDPR
that is being tracked is health, sex orientation and politics.
Degeling et al.~\cite{degeling2018we} quantified the changes of privacy policies on the Top-500 sites of the 28 EU countries.
They found that, in total, 85\% of the websites have a privacy policy, and that GDPR did not significantly change the way 3rd-party cookies are used.
In their most recent work, Sorenser et. al.~\cite{sorenser19} measured the changes on the presence of 3rd-parties, before and after GDPR enforcement, in 1200 popular websites across EU.
Their study shows that there were not significant changes in the general state of the web, and that GDPR had a potential effect only on specific types of websites.

\begin{figure*}[ht]
	\begin{subfigure}{.245\textwidth}
		\centering
		\includegraphics[width=1\columnwidth]{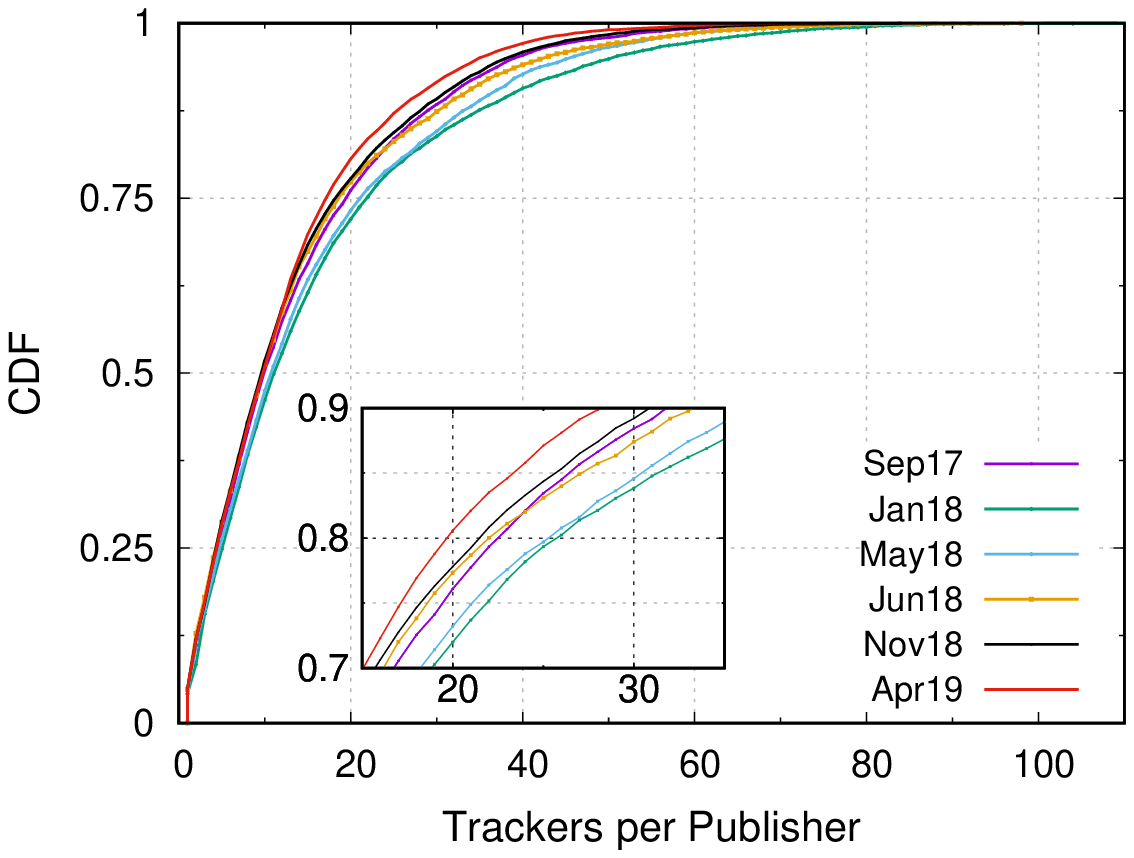}
		\subcaption{}
		\label{fig:cdf_unq}
	\end{subfigure}
	\begin{subfigure}{.245\textwidth}
		\centering
		\includegraphics[width=1\columnwidth]{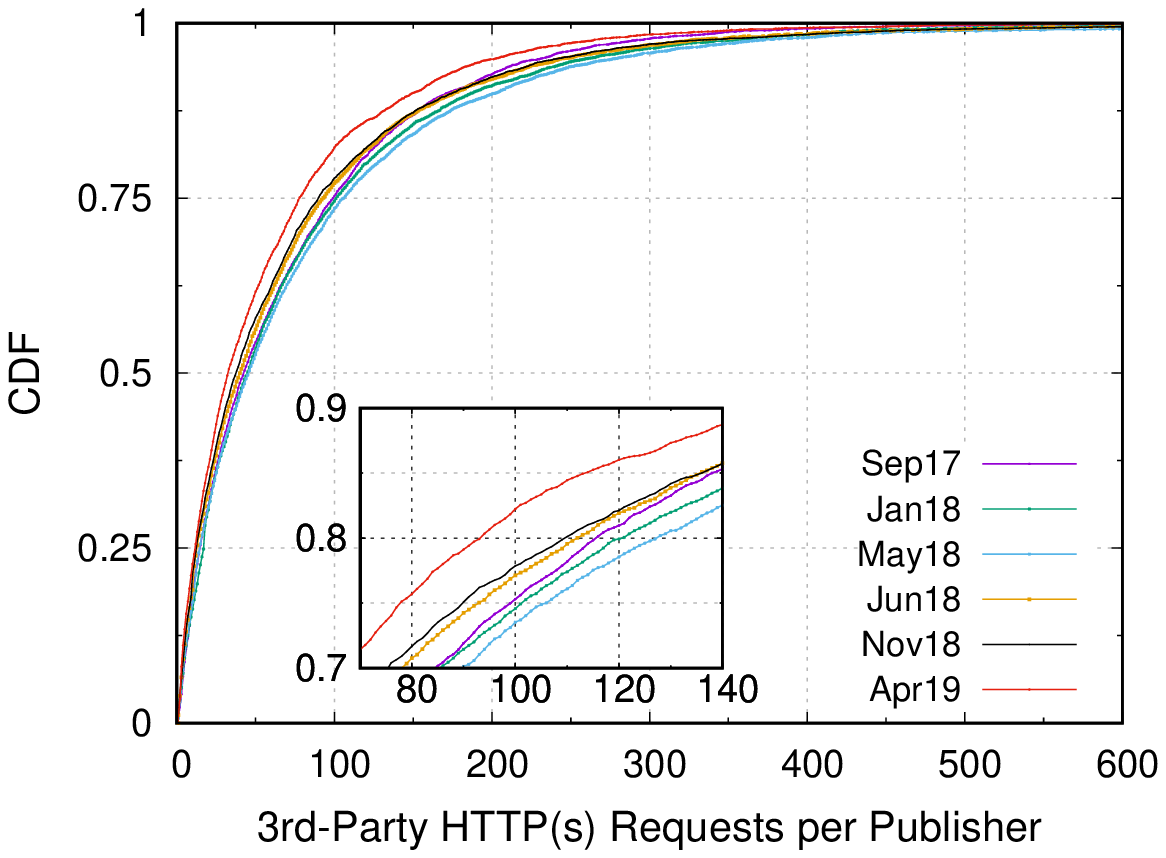}
		\subcaption{}
		\label{fig:cdf_total}
	\end{subfigure}
	\begin{subfigure}{.245\textwidth}
		\centering
		\includegraphics[width=1\columnwidth]{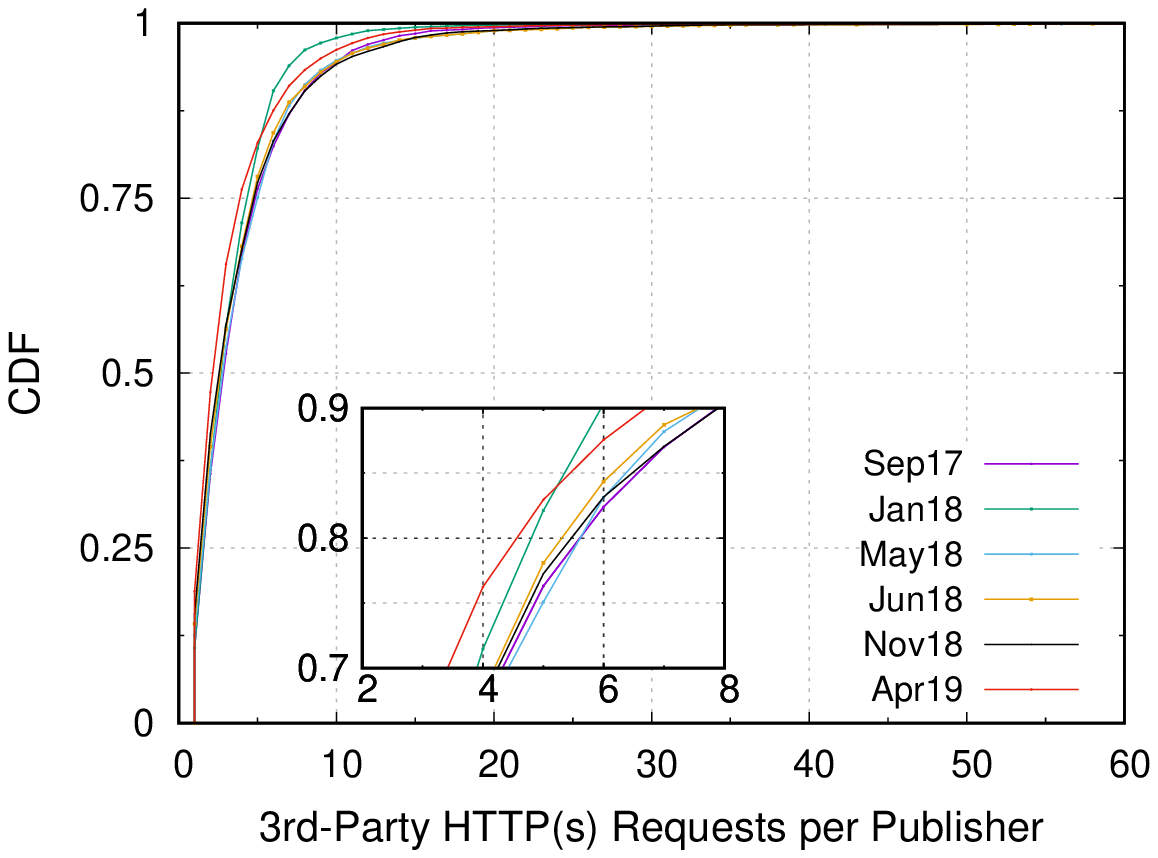}
		\subcaption{}
		\label{fig:cdf_avg}
	\end{subfigure}
	\begin{subfigure}{.245\textwidth}
		\centering
		\includegraphics[width=1\columnwidth]{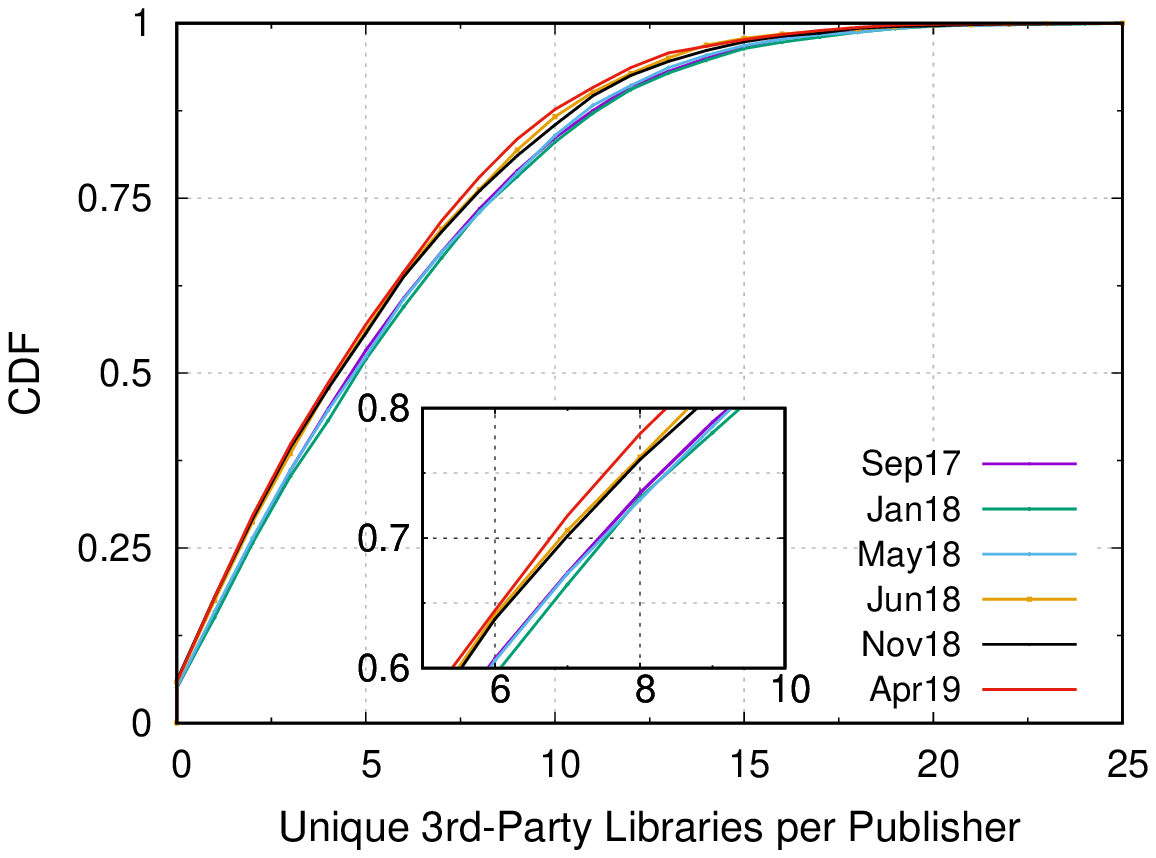}
		\subcaption{}
		\label{fig:cdf_post}
	\end{subfigure}
	\caption{These figures, from left to right, show (a) the number of unique 3rd-party trackers, (b) the total number of  3rd-party HTTP(s) requests, (c) the average number of 3rd-party HTTP(s) requests, and (d) the number of embedded 3rd-party libraries.}
	\label{fig:network}
\end{figure*}

	\section{Data Collection}
\label{sec:data_collection}
\begin{table}[t]
	\caption{Timeline of web crawls using top Alexa lists, and number of unique 1st- and 3rd-party domains we detected in each crawl. The (*) shows which snapshots were collected after the enforcement of GDPR. The overlap row indicates the intersection of common domains across all snapshots.}
	\label{tab:dataset}
	\begin{center}
		\resizebox{0.9\linewidth}{!}{  
			\begin{tabular}{c|cccc} 
				\toprule
				\textbf{Dataset} &\textbf{Alexa Ranks} & \textbf{1st-parties} &\textbf{3rd-parties}\\
				\midrule
				\rowcolor{Gray}
				September 2017 & 10K & 8311 & 848\\
				January 2018 & 30K &29444  &1036\\
				\rowcolor{Gray}
				May  2018 & 80K & 73493 &1096\\
				(*) June 2018 &80K & 73813&1068\\  
				\rowcolor{Gray}
				(*) November 2018 & 65K & 61287 &1002\\
				(*) April 2019 & 65K & 59662& 819\\
				\rowcolor{Gray}
				\textbf{Overlap}  &\textbf{10K} &\textbf{5100} &\textbf{527} \\
				\bottomrule
			\end{tabular}
		}
	\end{center}
\end{table}
	
Since our purpose is to conduct a longitudinal study on the evolution of tracking, and how it was affected by the recent EU regulations, we collected historical data covering a period of almost 2 years (September 2017 to April 2019).
For collecting these data we used the OpenWPM framework~\cite{englehardt_ccs} to crawl multiple websites via scripted browsers, and we stored all the HTTP(s) incoming and outgoing requests.
We also logged the cookies set by JavaScript, and stored various other crawl-related data (i.e., time of visit, HTML files, etc.).
During the crawling, we did not set the ``Do Not Track" flag, we configured our browser to accept all connections, and we empirically set the timeout for a website to respond to 30 seconds.
For not introducing any unnecessary complexity and overhead during the crawls, we do not perform any browsing to visit deeper than the main page of each website.
We deployed the framework on a single computer at an EU academic institution, having a unique IP address, thus, avoiding any content biases or any type of location based discrimination.
As a website input corpus, we used the Alexa Top 100K list~\cite{alexa}, and based on our available resources, we crawled each time the top websites covering different subsets of the list.
Starting from September 2017, we repeated our crawls approximately every 5 to 6 months.
We also performed two consecutive crawls just before and after the GDPR enforcement, i.e., mid May 2018 and beginning of June 2018.

In each snapshot of our dataset, we define two different entities:
(i) \textit{Publishers}, which are the websites that the users explicitly visit (\textit{i.e., 1st-parties}), and (ii) \textit{Trackers}, which refer to the \textit{3rd-party} domains that are embedded within the visited pages (i.e., domains from which resources are fetched, that set cookies, serve content, etc.).
We use the most recent, by the time of each crawling, Disconnect List~\cite{disco} (a popular browser ad-blocking list), in order to identify which requests are directed towards trackers.
In this way, we were able to accurately detect the presence of trackers in each page and to identify their behaviors.
Since tracking domains change frequently, and only a subset of them might remain constant through time, the list is constantly updated to capture such changes.
Furthermore, we only consider publishers that embed at least one tracker in their page.
For simplicity, we interchangeably refer to 1st-parties or publishers, and to 3rd-parties or trackers in the rest of the paper.
A detailed description of our datasets is presented in Table~\ref{tab:dataset}.

	\section{Network Activity of Trackers}
\label{sec:tracking_activity}
	
Using the aforementioned six time snapshots, we perform an analysis on the network activity of trackers that are embedded in the publishers crawled.
As shown in Table~\ref{tab:dataset}, the snapshots in our datasets include different subsets of the top websites. 
Also, considering that the Alexa list is volatile and that it is constantly changing~\cite{scheitle2018long}, we focus our analysis on a subset 
of 5100 publishers that are common across all time snapshots.

\noindent
\textbf{HTTPS Adoption}.
Before digging deeper into the patterns and the effects of the online tracking ecosystem, here we make a preliminary measurement of the HTTPS adoption by the publishers.
In our first snapshot (i.e., September 2017) we found that 3144 (61.64\%) of the publishers were using HTTPS.
This percentage is reasonable considering that our corpus was formed by publishers that rank among the Top-10K of the 
Alexa list, and that those tend to be more regulation compliant and probably privacy-preserving, as also reported in previous work~\cite{felt2017measuring}.
In the snapshot of January 2018, we found 944 additional publishers (80.15\% of our dataset) that adopted HTTPS, while $\sim$92\% of the publishers adopted it by April 2019, validating the shift towards more secure practices.

\noindent
\textbf{Tracking Activity}.
Figure~\ref{fig:network} summarizes different dimensions of the tracking activity of 3rd-parties that are present in the publishers of our datasets.
The unique number of trackers embedded in each publisher (i.e., based on the outgoing HTTP requests observed) is shown in
Figure~\ref{fig:network}(\subref{fig:cdf_unq}).
We observe that, in general, there is a clear decrease in the number of trackers through time. Specifically, until January 2018, 50\% of the publishers communicated with almost 10 individual trackers in every visit, but this changed by the end of April 2019, when this number dropped to almost half, reaching the lowest measured value.
This might not be a typical trend of the  ecosystem, and this ``momentary'' elimination of trackers could be caused by the enforcement of GDPR in May 2018, but further investigation is needed to see whether publishers will go back to the same levels of tracking as before GDPR.

In Figure~\ref{fig:network}(\subref{fig:cdf_total}) we report the \emph{total} number of 3rd-party requests per publisher.
We observe that the trend on the distributions is similar: through time there is a decline in the average number of 3rd-party requests per publisher. 
Summarizing the previous two measures, in Figure~\ref{fig:network}(\subref{fig:cdf_avg}) we show the \emph{average} number of 3rd-party requests per publisher.
In theory, since both the number of trackers and the total requests are in decline, their ratio (i.e., average) should also be reduced through time.
This trend is observed for the majority of websites, but for a small set of the publishers in the last snapshot ($\sim$10\%), this number increases.

Since 3rd-party requests may be issued for different reasons (e.g., cookie delivery, tracking pixels, content, ad-libraries), in Figure~\ref{fig:network}(\subref{fig:cdf_post}), we report the unique number of trackers that deliver 3rd-party libraries (i.e., JS) to publishers.
For this measurement, we consider that connections to different subdomains, e.g.,  \textit{subdomain1.domain.com/ad-library.js} and \textit{subdomain2.domain.com/ad-library.js}, result in a connection to a unique 3rd-party provider.
From the results, we note that by 2017, up to 50\% of the publishers communicated with at least 5 different providers, and by 2019 this number decreased to 4.
Also, there is a reduction in the plurality of the embedded  providers (e.g., Amazon and Tapad were not present in the latest snapshots), which potentially results in higher tracking activity by the remaining entities, or increases the chances of cookie sharing and synchronization.
Previous works on this topic (e.g.,~\cite{iordanou2018tracing,urban2018unwanted,sorensen2019before}) also report similar decline in the number and frequency of trackers.
Specifically, in~\cite{sorensen2019before} they reported that fewer 3rd-parties are present in specific categories of websites.
These facts might be a side effect of the GDPR being enforced, or other effects of the general evolution of the Web tracking ecosystem.

	\section{Publishers \& Trackers: The PT Graph}
\label{sec:full-bigraphs}

\noindent
\textbf{Graph Construction}.
\noindent
We follow a similar approach to Kalavri et al.'s~\cite{kalavri2016like} for constructing our graphs. We create a set of \textit{2-mode} graphs of the publishers and their associated trackers, where the edges of each graph connect vertices of different modes. In this graph, a publisher can connect to multiple trackers, and a tracker can connect to multiple publishers.

We represent all the domains that a browser requests as a 2-mode graph, by creating a set of mappings.
The \textbf{$V_P$} represents the set of websites (publishers) a user visits, and accordingly \textbf{$V_T$} is  the set of trackers embedded in publishers.
We also define \textbf{$E_w$} as the set of weighted edges connecting vertices of the two different modes, and \textbf{$w=(i,j)$} as the weight of the edge connecting tracker $i$ with publisher $j$.
Furthermore, we go beyond the state-of-art (i.e.,~\cite{kalavri2016like}) and add weights on the edges, to represent the number of HTTP(s) requests between a publisher and a tracker. The weight $w$=$(i,j)$, encodes the number of times that a tracker \textit{i} communicated via HTTP requests with a publisher \textit{j}.

\noindent
\textbf{Data Filtering \& Graph Metrics.}
As reported in Section~\ref{sec:data_collection}, in each snapshot, we crawled a subset of the top Alexa list. 
Since we want to create a connected representation of the bipartite graphs, we use the Largest Connected Component (LCC) of each graph.
We found that there are some isolated groups of nodes, that include websites that communicated with one or two different, but not popular trackers; we exclude such isolated groups from the graphs. 
In each of our final graphs, the LCCs contain on average $\sim$99\% of the publishers and $\sim$95\% of trackers of the originally crawled lists.

This type of connected graphs allows us to apply various graph metrics to quantify their properties, and compare them across time.
In our analysis, we use graph metrics similar to those of previous works on this topic (e.g.,~\cite{kalavri2016like,bashir2018diffusion,urban2018unwanted}).
Specifically  we compute the Density and Diameter to study the inner structure of each network and the connectivity between their nodes.  
Other metrics like the Average Clustering Coefficient, Degree Centrality and  Betweenness Centrality reveal the properties of each node in terms of the shortest paths that pass through the specific edge, and the number of connected neighbors.
Finally, we also use Coreness-Periphery, a metric not studied before in the literature, that measures the importance of each node in terms of how ``involved or core" it is inside the network.
Detailed definitions of these metrics can be found in~\cite{freeman1977set,girvan2002community,rombach2017core}.
The generated graph models and their properties for each snapshots are given in Table~\ref{tab:union_graphs}.

\subsection{Stability of PT graph properties over time}

The differences in the number of nodes and edges between these graphs reflect the different number of visited publishers in each snapshot.
In general, the average clustering coefficient measures the degree with which the nodes of a graph tend to cluster together (i.e., tend to close triangles between triplets of nodes, or quadruples in bipartite graphs).
The low value of this metric on each of the graphs is associated with the bipartite connectivity between the sets of nodes~\cite{latapy2006basic}.
This metric, in conjunction with the low density, reveals sparse connections between the different groups of nodes.

Regarding the number of trackers, there is a stable reduction through time, matching the measured decrease in the average number of trackers, as discussed in Section~\ref{sec:tracking_activity}.
In general, the characteristics and distance metrics computed on the graphs reveal a consistent structure of the ecosystem during the focal period of our analysis, except for the most recent snapshot where the diameter reaches its lowest value, while the number of edges increases.
This trend captures an increase in the connectivity of the graph nodes, pointing to tracker nodes being closer in the graph, as also discussed next.
\begin{table}[t!]
\begin{center}
 \caption{PT graph characteristics: number of vertices (N);  edges (E); normalized average weight per edge (W); average clustering coefficient (CC); density (DE); diameter (D).}
 \label{tab:union_graphs}
\resizebox{0.9\linewidth}{!}{  
\begin{tabular}{c|cccccc}
	\toprule
	\textbf{Dataset} & \textbf{N} &  \textbf{E} & \textbf{W}& \textbf{CC} & \textbf{DE} &  \textbf{D} \\
	\midrule
	\rowcolor{Gray}
Sep17 & 5710 & 74037 &0.0013&  0.024 & 0.022 & 7 \\
 Jan18 & 5688 & 86875 &0.0012&  0.024& 0.027& 7 \\
  \rowcolor{Gray}
 May18 & 5678 & 81717 &0.0013&  0.021 &   0.026& 8 \\
 Jun18 & 5654 & 76077 &0.0013&  0.023 &  0.025&7 \\
  \rowcolor{Gray}
 Nov18 & 5636 & 71481&0.0013&  0.026& 0.024 & 8  \\
  Apr19 & 5602 & 72722 &0.0011& 0.022 & 0.025&4 \\
 \bottomrule
\end{tabular} 
}
\end{center}
\end{table}

\begin{figure*}[ht]
	\begin{subfigure}{.330\textwidth}
		\centering
		\includegraphics[width=1\columnwidth]{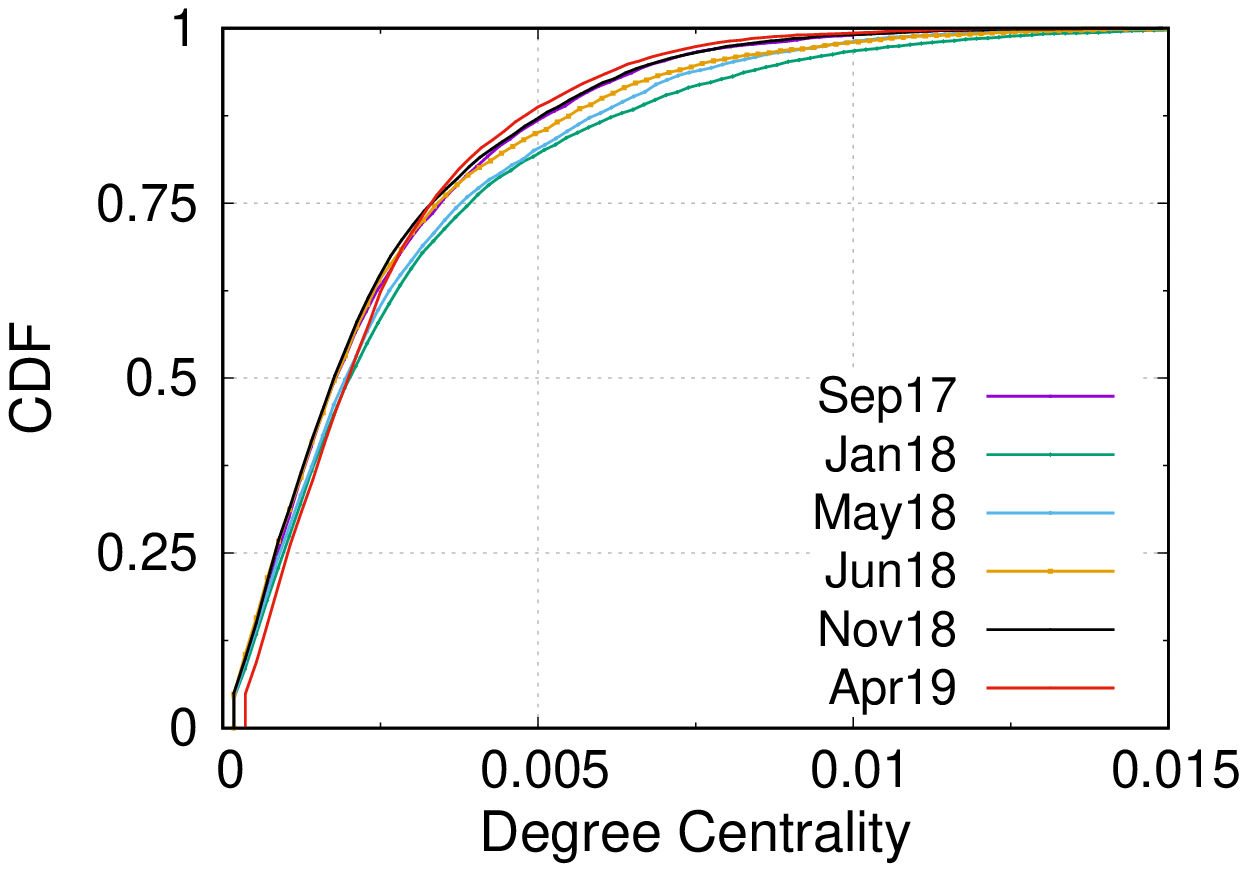}
		\subcaption{}
		\label{fig:cdf_degree_publishers}
	\end{subfigure}
	\begin{subfigure}{.330\textwidth}
		\centering
		\includegraphics[width=1\columnwidth]{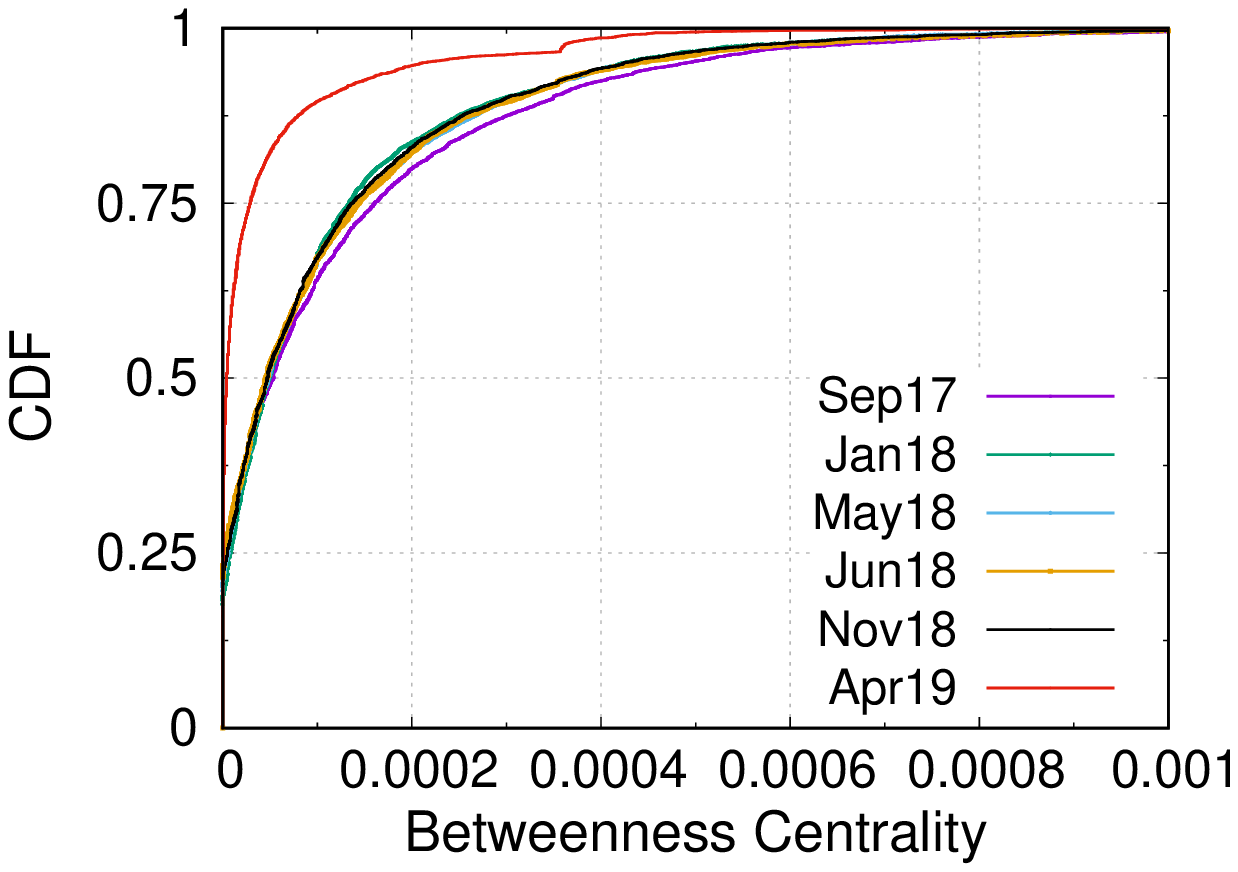}
		\subcaption{}
		\label{fig:cdf_betweenness_publishers}
	\end{subfigure}
	\begin{subfigure}{.330\textwidth}
		\centering
		\includegraphics[width=1\columnwidth]{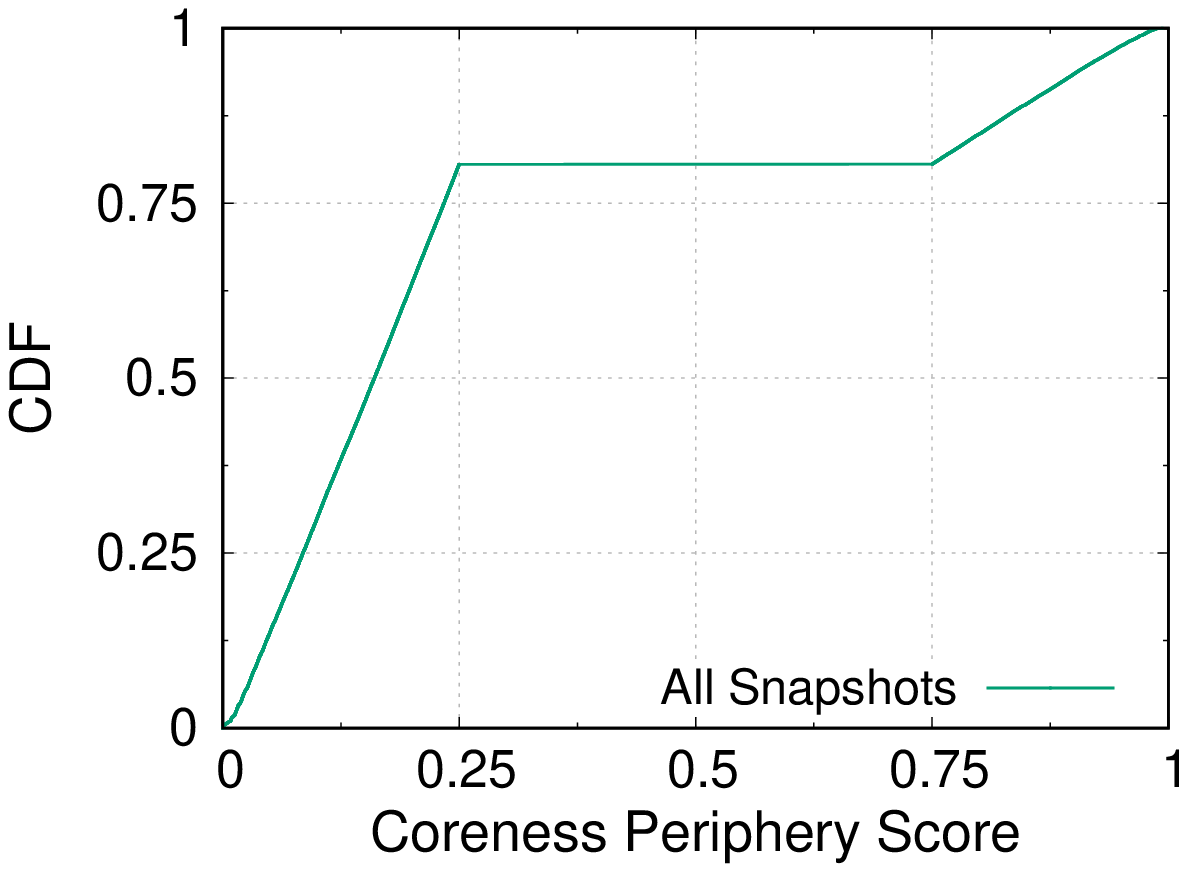}
		\subcaption{}
		\label{fig:cdf_coreness_publishers}
	\end{subfigure}
	\begin{subfigure}{.330\textwidth}
		\centering
		\includegraphics[width=1\columnwidth]{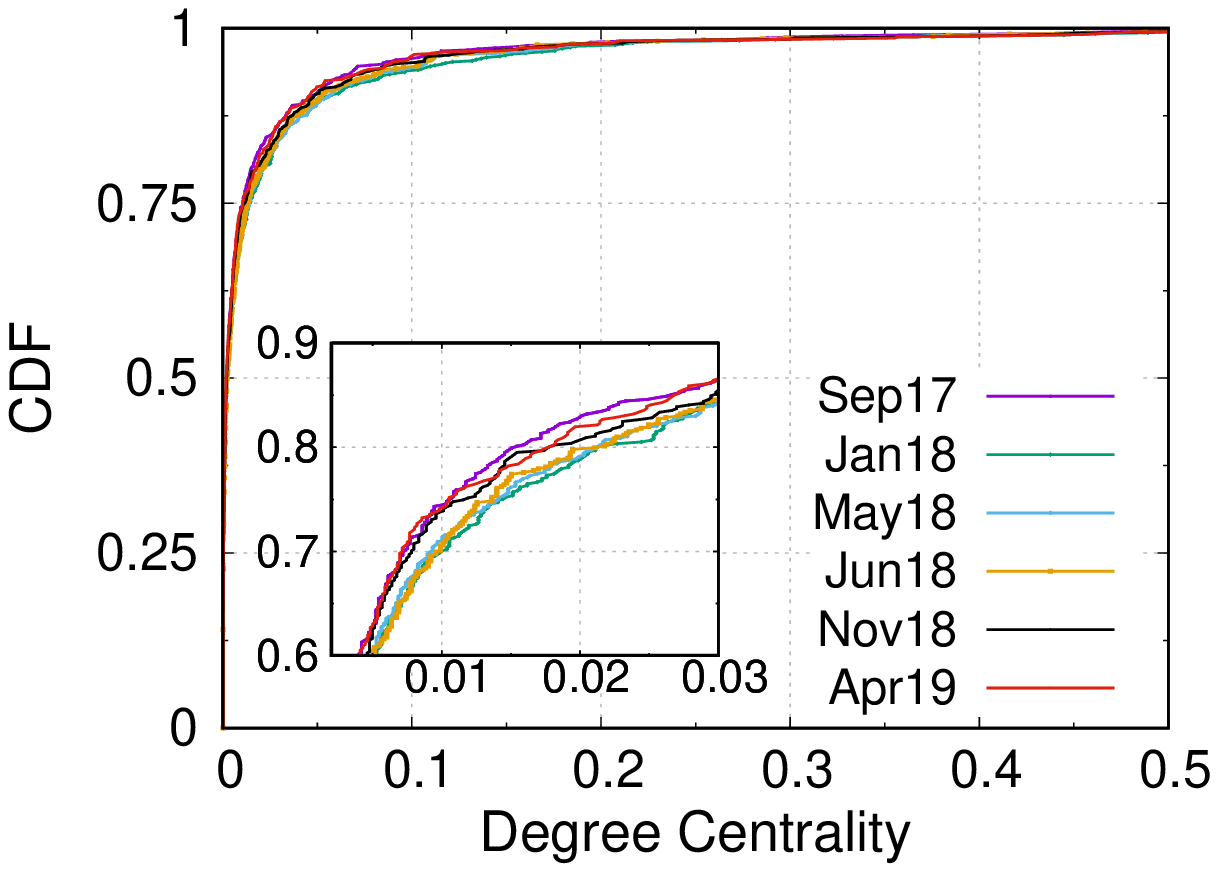}
		\subcaption{}
		\label{fig:cdf_degree_trackers}
	\end{subfigure}
	\begin{subfigure}{.330\textwidth}
		\centering
		\includegraphics[width=1\columnwidth]{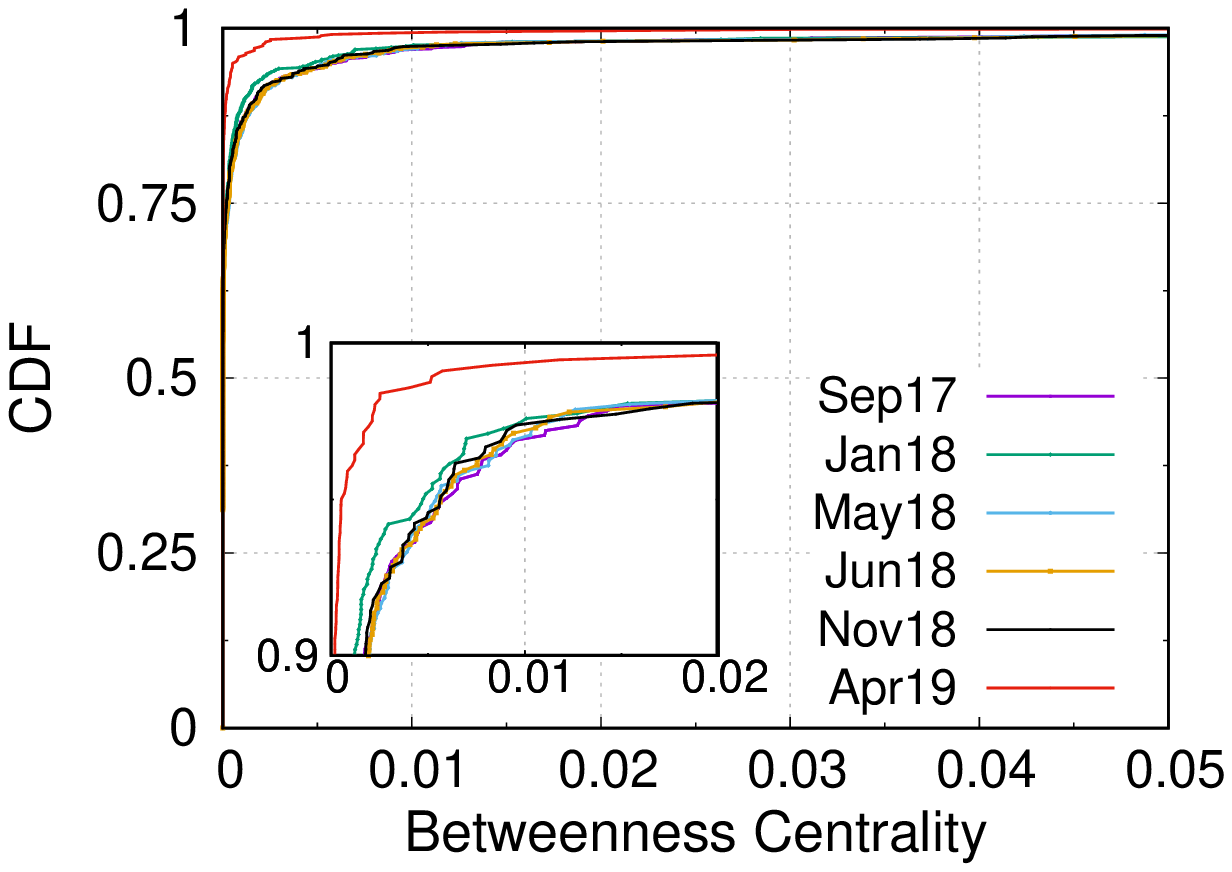}
		\subcaption{}
		\label{fig:cdf_betweenness_trackers}
	\end{subfigure}
		\begin{subfigure}{.330\textwidth}
		\centering
		\includegraphics[width=1\columnwidth]{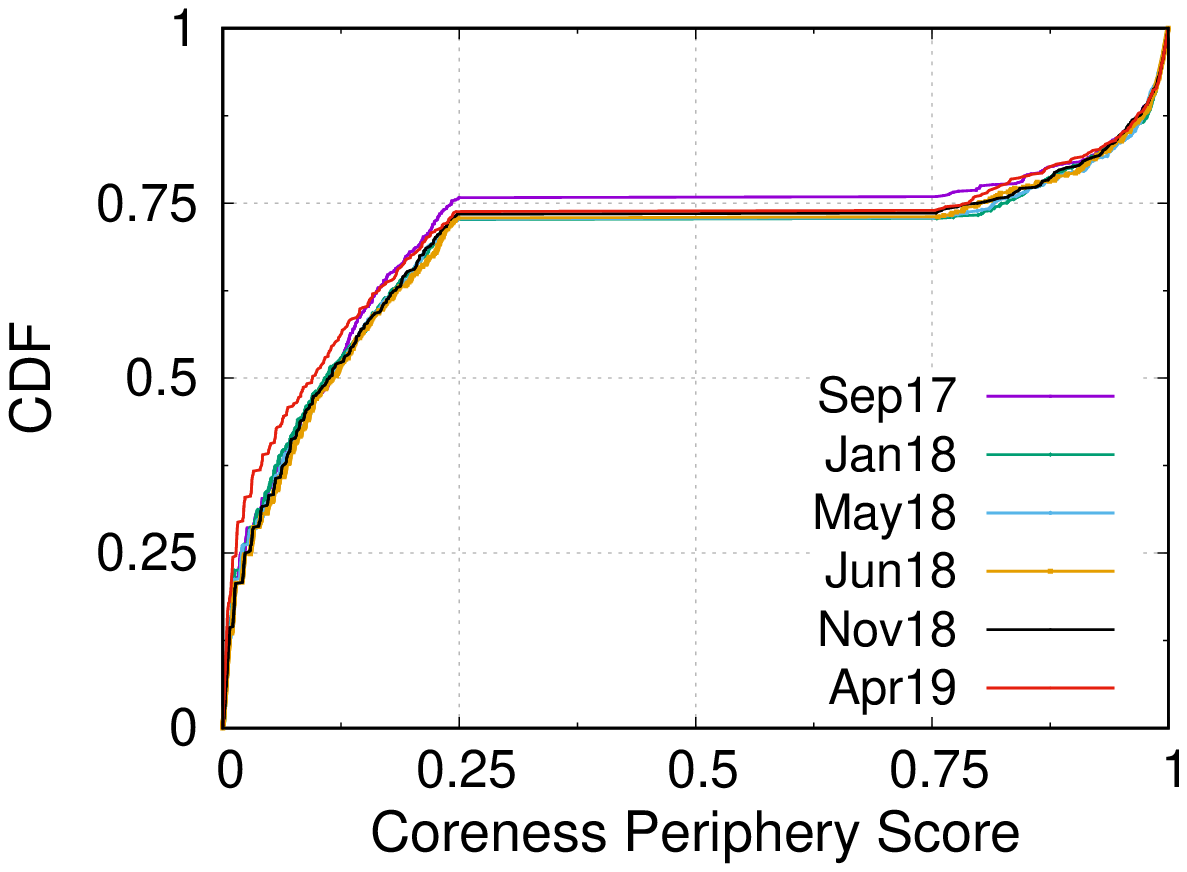}
		\subcaption{}
		\label{fig:cdf_coreness_trackers}
	\end{subfigure}
	\caption{Top row shows metrics for Publishers and bottom row for Trackers: (a) and (d) Normalized degree centrality; (b) and (e) Betweenness Centrality; (c) and (f) Coreness periphery for each PT Graph.}
	\label{fig:metrics}

\end{figure*}

\noindent
\textbf{Stability of Publishers vs. Trackers}.
We measure the degree, betweenness centrality and coreness-periphery of each node in our bipartite graphs, in order to quantify the backbone structure of each graph in terms of connectivity and centrality of nodes.
Figure~\ref{fig:metrics}, plots the distribution for each metric, as computed on each snapshot for publishers and trackers.

Focusing our analysis on Figures~\ref{fig:cdf_degree_publishers} and~\ref{fig:cdf_degree_trackers}, the degree centrality for publishers is an order of magnitude lower than the one for trackers.
Also, 50\% of the publishers' degree centrality is $\leq$ 0.025, with only $\sim$ 5\% of the publishers having more than 0.01.
Conversely, tracker nodes have higher degree centrality scores, with $\sim$ 5\% measured at $\geq$0.15, hinting to the fact of the well-known trackers that cover approximately all the publishers (e.g., Google and Facebook).
Some examples of publishers with high degree centrality are \textit{telegraph.co.uk}, \textit{newyorker.com} and \textit{rollingstone.com}, and examples of trackers with high degree centrality include \textit{google-analytics.com}, \textit{criteo.com}, and \textit{facebook.com}.
By definition, since we are analyzing a bipartite network constructed by two sets of nodes with different sizes and edge weights, it is reasonable for the tracking nodes to be more central in the network structure.

The betweenness centrality for publishers and trackers
are given in Figures~\ref{fig:cdf_betweenness_publishers} and~\ref{fig:cdf_betweenness_trackers}, respectively.
Publishers' low scores are expected, as betweenness centrality measures the extent to which a node lies on paths between other nodes. 
Obviously, publishers are not connected to each other but only with trackers.
Trackers that are highly active, central and well known, are found on the tail of the distribution, with scores $\geq$ 0.02.
Example of publishers with high betweenness are \textit{starbucks.com}, \textit{livescore.com} and \textit{cnn.com}, and accordingly trackers with high betweenness are \textit{moatads.com}, \textit{instagram.com} and \textit{scorecardresearch.com}.

Finally, when comparing coreness scores for publishers and trackers, 
in Figures~\ref{fig:cdf_coreness_publishers} and~\ref{fig:cdf_coreness_trackers}, respectively, trackers tend to occupy positions in the graph with higher coreness than publishers.
Interestingly, we observe two classes of nodes in each graph, with periphery nodes being in the beginning of each CDF (up to $\sim$0.25 coreness score), and core nodes being in the end of each CDF, with coreness score $\geq$0.75.
Well-known publishers such as \textit{sfgate.com}, \textit{sport.es} and \textit{indianexpress.com}, and trackers such as \textit{yandex.ru}, \textit{adroll.com} and \textit{amazon-adsystem.com}. have high coreness-periphery scores.

\subsection{How do these centrality metrics correlate?}

After studying the inner structure of the bipartite networks, we evaluate the relationship between the centrality of nodes for these three metrics.
This study will help us understand if the nodes (trackers or publishers) tend to be top (or bottom) in all metrics at the same time, or if there is some disassociation between these metrics.
Such disassociation can reveal outliers of nodes who are high in one metric but low in another.

For this reason, we compute the Pearson Correlation score between the distributions of Degree Centrality, Betweenness Centrality  and Coreness-Periphery for the trackers and publishers, independently.
A detailed report on the correlation scores is given in Table~\ref{tab:correlation}.
In general, we find a strong association between all the distributions with high confidence level.
Trackers have higher correlation score in the Degree-Betweenness comparison, hinting their importance and central role on the network, regardless of the metric used.
On the contrary, the correlation scores for Degree-Coreness, for the publishers, are measured higher than trackers, again validating the importance of the publishers who tend to have high and important position in the network structure.
Finally, across both types of nodes, Betweenness-Coreness correlation scores are lower, pointing to a disassociation between the two measures.

\begin{table}[t]
\caption{Pearson Correlation Coefficient between the distribution of: Degree  \& Betweenness Centrality (\textit{DC-BC}), Degree Centrality \& Coreness Periphery (\textit{DC-CP}), and Betweenness Centrality \& Coreness Periphery (\textit{BC-CP}), for each set of the PT graph for common publishers across snapshots.
All results reported are statistically significant at $p-value$$<$$0.009$.
The publishers are ranked according to the Alexa list, while trackers according to their Degree Centrality.}
\label{tab:correlation}
\resizebox{1.0\columnwidth}{!}{  
\begin{tabular}{c|ccc|ccc}
\toprule
    \multirow{2}{*}{\textbf{Dataset}} &
      \multicolumn{3}{c}{\textbf{Publishers}} &
       \multicolumn{3}{c}{\textbf{Trackers}} \\
    & \textit{$\sim$DC-BC} & \textit{$\sim$DC-CP}& \textit{$\sim$BC-CP} & \textit{$\sim$DC-BC} & \textit{$\sim$DC-CP}& \textit{$\sim$BC-CP}  \\
    \midrule
    \rowcolor{Gray}
    Sep17 &  0.50 & 0.69 &0.26&  0.75 & 0.47 &0.26  \\
    Jan18 & 0.43	 & 0.83 	 &  0.3&0.73	 & 0.50 	 & 0.24\\
\rowcolor{Gray}
    May18 & 0.47	 & 0.75	 &  0.20 & 0.73	 & 0.50 	 & 0.24\\
   Jun18 & 0.46 	 & 0.74 	 &  0.24 &0.75	 & 0.48	 & 0.25\\
\rowcolor{Gray}
  	Nov18 & 0.42 	 & 0.72 	 &  0.19 &0.77 	 & 0.49 	 &  0.26\\
    Apr19 &  0.43 	 & 0.71 	 &  0.23 &0.54 	 & 0.47 	 &  0.18\\
    \bottomrule
  \end{tabular}}
\end{table}

\subsection{What is the ecosystem's current state?}

\begin{table}[htp!]
\caption{Top-25 Trackers ranked by Degree Centrality and labeled under the umbrella of company/organization and the average percentage of coverage in publishers through time.\\
(*): Trackers who were Top-25 trackers across all snapshots.\\
(+/-): Trackers who were Top-25 in one or more snapshots, but their rank decreased through time below Top-25.\\
(+): Set of trackers that were not part of the Top-25 in the first snapshot, and they climbed in the Top-25 through time.}
\label{tab:top_trackers_degree}
\begin{center}
\resizebox{1.0\linewidth}{!}{  
\begin{tabular}{|l||l|l|}
\hline
 \textbf{Tracker} & \textbf{Organization} & \textbf{Publishers(\%)}\\
 \hline
(*) google-analytics.com & Google& 81.0 \\
(*) doubleclick.net&  &  70.0\\
(*) google.com &  & 51.0 \\
(*) googleapis.com &  &  57.5\\
(*) googletagmanager.com &  & 36.5 \\
 \hline
(*) facebook.com & Facebook & 44.5 \\
(*) facebook.net &  &  41.5\\
 \hline
(*) googletagservices.com& Google&  28.0\\
(*) gstatic.com& & 44 \\
(*) googlesyndication.com & & 28.3 \\
(*) googleadservices.com & & 19.0 \\
 \hline
 (*)  cloudfront.net & Amazon & 18.0 \\
\hline
 (*)  adnxs.com & App Nexus& 18.0 \\
\hline
 (*)  criteo.com &  & 13.0 \\
 (*) criteo.net & Criteo& 13 \\
\hline
(*) scorecardresearch.com & comScore& 12.5 \\
\hline
(*) twitter.com & Twitter& 20.0 \\
\hline
(*) rubiconproject.com & Google& 12.5 \\
\hline
(*) pubmatic.com & Pubmatic&  11\\
\hline
(*) openx.net& OpenX&  8.5\\
\hline
(*) casalemedia.com & Casale Media & 9.0 \\
\hline
(*) advertising.com & Verizon Media & 7.0\\
\hline
\hline
\rowcolor{Gray}
(+/-) quantserve.com& Quantcast& 9.0 \\
\hline
\rowcolor{Gray}
(+/-) adsrvr.org & The Trade Desk  & 9.0 \\
\hline
\rowcolor{Gray}
(+/-) taboola.com & Taboola, Inc & 7.0 \\
\hline
\rowcolor{Gray}
(+/-) nr-data.net & New Relic & 8.0 \\
\hline
\rowcolor{Gray}
(+/-) 2mdn.net & Google & 6.0 \\
\hline
\rowcolor{Gray}
(+/-) bluekai.com & BlueKai & 8.0 \\
\hline
\hline
\rowcolor{ashgrey}
(+) alexametrics.com& Amazon& 4.0 \\
\hline
\rowcolor{ashgrey}
(+) demdex.net& Adobe& 7.0 \\
\hline
\rowcolor{ashgrey}
(+) newrelic.com & New Relic & 8.0  \\
\hline
\end{tabular} 
}
\end{center}
\end{table}

After analyzing the graphs' inner properties, we want to quantify the importance of the top tracking nodes for the ecosystem and their impact to user privacy.
To have a clear view of the most important trackers at each time of our crawl, we measured their degree centrality and ranked them accordingly. 
A complete report about the Top-25 trackers across time is given in Table~\ref{tab:top_trackers_degree}.
Interestingly, the Top-22 trackers remained the same across all snapshots with minor fluctuations on the internal ranking.
The most important trackers contain the ``big", well-known entities of the ad-industry, such us \textit{Google}, \textit{Facebook}, \textit{Twitter} and \textit{Criteo}, as well as some smaller, but well established such as \textit{Bluekai} and \textit{Taboola}.
This table also reports those trackers that gradually climbed into the Top list, illustrating the plurality of the tracking ecosystem.

Furthermore, in Table~\ref{tab:top_trackers_betweenness}, we make a similar investigation for trackers' betweenness centrality.
We note that the top of the list is populated by similar trackers.
However, new entities such as \textit{linkedin.com}, \textit{ads-twitter.com} and \textit{everesttech.net} emerge, demonstrating a central position in the ecosystem with respect to mediating flows between distant parts in the ecosystem.

Overall, the almost immutable list of top trackers in either of the two metrics points to the fact that the GDPR enforcement had no effect on them, either in their importance in the web tracking ecosystem, or their coverage across websites.
Also, in the previous sections, we found that trackers are present in more websites (as time passed by), but at the same time, web requests have been reduced.
We can conclude that there may be a ``shift" of publishers on the type of business relationships they make with the well-known and GDPR-compliant trackers.

\begin{table}[t]
\caption{Top-25 Trackers ranked by Betweenness Centrality (BC) and labeled under the umbrella of company/organization, with their BC score and coverage in publishers by April 2019.
We highlight the trackers not present in Table~\ref{tab:top_trackers_degree}.}
\label{tab:top_trackers_betweenness}
\begin{center}
\resizebox{1.0\linewidth}{!}{  
\begin{tabular}{|l||l|l|l|}
\hline
 \textbf{Tracker} & \textbf{Organization}& \textbf{BC} & \textbf{\% Publishers}\\
 \hline
googletagmanager.com  &  Google & 0.077     & 46.55    \\
doubleclick.net  &  & 0.012     & 72.15    \\
googleadservices.com  &   & 0.008     & 19.59    \\
googletagservices.com  &  & 0.006     & 28.82    \\
gstatic.com  &  & 0.005     & 44.75    \\
\hline
cloudfront.net  & Amazon  & 0.005     & 14.52    \\
\hline

newrelic.com  & New Relic  & 0.004     & 7.63   \\
\hline
\rowcolor{ashgrey}

rlcdn.com  & Live Ramp & 0.003     & 8.66   \\
\hline

pubmatic.com  &  Pubmatic & 0.003   & 13.1   \\
\hline

google.com  & Google & 0.030   & 59.09    \\
\hline
\rowcolor{ashgrey}

nr-data.net  & New Relic & 0.002     & 7.61   \\
\hline

facebook.com  &  Facebook & 0.002     & 44.01    \\
\hline
\rowcolor{ashgrey}
everesttech.net  & Adobe & 0.002     & 3.52   \\
\hline

casalemedia.com  & Casale Media & 0.002     & 7.48   \\
\hline

alexametrics.com  &  Amazon & 0.002     & 4.46   \\
\hline
\rowcolor{ashgrey}
ads-twitter.com  & Twitter & 0.002     & 4.72   \\
\hline

adsrvr.org  &   Trade Desk& 0.002     & 5.67   \\
\hline

adnxs.com  & App Nexus & 0.002     & 13.83    \\
\hline

twitter.com  & Twitter & 0.001     & 13.32    \\
\hline

rubiconproject.com  & Google  & 0.001     & 10.15    \\
\hline
\rowcolor{ashgrey}
quantcount.com  & Quantcast  & 0.001     & 4.11   \\
\hline
openx.net  &  OpenX & 0.001     & 8.59   \\
\hline
\rowcolor{ashgrey}
linkedin.com  & Microsoft  & 0.001     & 4.71   \\
\hline
advertising.com  & Verizon Media & 0.001     & 7.23   \\
\hline

\end{tabular} 
}
\end{center}
\end{table}

\section{Tracker to Tracker: The TT Graph}
\label{sec:tt_graphs}

\noindent
\textbf{Graph Construction.}
\noindent
We build tracker-to-tracker graphs that are undirected but weighted, \textit{$TT= (V_{TT}, E_{TT})$}, and originate from their corresponding PT graphs.
Similarly to Section~\ref{sec:full-bigraphs}, we create different sets of nodes that have specific properties.
In our TT graphs, \textbf{$V_{TT}$} represents the set of trackers embedded in publishers, and \textbf{$E_{TT}$} is the set of weighted edges connecting two trackers, if and only if both trackers coexist in at least two different publishers.
Also, the weight \textit{w=(i,j)}, encodes the number of publishers that tracker \textit{i} and \textit{j} coexisted.
A detailed description of our $TT$ graphs is presented in Table~\ref{tab:tt_graphs}.

The $TT$ graph of each snapshot has a fairly dense structure (average Density $0.19-0.21$).
The number of nodes, as well as the number of edges, are comparable through snapshots, which is reasonable since we focused only on the common publishers in the dataset and extracted the trackers present in each snapshot.
Interestingly, in May 2018 (before GDPR) the number of edges reached a maximum which was not surpassed in subsequent snapshots.
In general, in all $TT$ graphs, trackers are well connected and clustered with each other (average clustering coefficient $0.67-0.72$).
These properties of the $TT$ graphs highlight the dense structure of the tracking ecosystem, and how 3rd-parties potentially share user's information.

\begin{table}[t]
\begin{center}
\caption{Characteristics of TT graphs produced from the PT graphs with common publishers across all snapshots.}
\label{tab:tt_graphs}
\resizebox{0.9\columnwidth}{!}{  
\begin{tabular}{c|cccccc}
\toprule
\textbf{Dataset} & \textbf{N} &  \textbf{E} & \textbf{W}& \textbf{CC} & \textbf{DE} &  \textbf{DD} \\
\midrule
\rowcolor{Gray}

Sep17 & 815 & 63177 &   0.004& 0.69  &0.19 &  4 \\

Jan18 & 774 & 53325 &  0.006&  0.67 & 0.17& 4 \\
\rowcolor{Gray}
May18 &  846 & 76686 &  0.005& 0.72 &   0.21& 4 \\

Jun18 & 824 & 69145 &  0.005&  0.70 &  0.20 &4 \\
\rowcolor{Gray}

Nov18 & 834 & 72265&  0.005& 0.71& 0.20 & 4  \\

Apr19 & 841 & 74012 &  0.005& 0.71 & 0.20 &4 \\
\bottomrule
\end{tabular}
}
\end{center}

\end{table}
\subsection{Are cookie synchronization pairs present in TT graphs?}

Since the purpose of the $TT$ graphs is to study potential data sharing among trackers, we need to compare the constructed pairs with existing data that already measure data sharing flows (i.e., ground truth datasets).
We received access to two CS datasets provided by Papadopoulos et. al.~\cite{papadopoulos2019cookie} and Bashir et al.~\cite{bashir2018diffusion}.
These datasets contain pairs of 3rd-parties performing CS while real users~\cite{papadopoulos2019cookie}, or crawlers~\cite{bashir2018diffusion} browsed the web.

Interestingly, the dataset from~\cite{papadopoulos2019cookie} also includes a normalized frequency on each pair, encoding the number of times the two entities of the pair shared information.
Following a similar representation as with the $TT$ graphs, we create 2 undirected CS graphs, $CS$=$(V_{CS}, E_{CS})$, with weighted edges for the data from~\cite{papadopoulos2019cookie}, i.e., \textit{w(i,j)} is the number of times that a pair of trackers $(i,j)$ performed information exchange.
The first CS graph from~\cite{papadopoulos2019cookie} has 4656 trackers and 8582 edges connecting them, whereas the second CS graph from~\cite{bashir2018diffusion} has 59 trackers and 200 edges connecting them.
To investigate the existence of CS pairs into $TT$ pairs, we define the following sets:
\begin{itemize}
\item $E_{CS}$ for the set of edges in a CS graph; $E_{TT}$ for the set of edges in a $TT$ graph.
\item $\neg CS$ for  the set of non-edges in a CS graph;  $\neg TT$ the set of non-edges in a $TT$ graph.
\end{itemize}
We also define the following overlaps of the above sets:  $O_{common}$ = $E_{CS} \cap E_{TT}$,$O_{\neg CS}$ = $\neg CS  \cap E_{TT}$ and $O_{\neg TT}$ = $ E_{CS} \cap \neg TT$.
To have an accurate measurement between the different overlaps, we filter the edges and store only those that are part of the common trackers between each CS and each $TT$ graph.
A detailed report on the percentages of overlap between each of the two CS graphs is given in Tables~\ref{tab:cs_pairs_tef} and~\ref{tab:cs_pairs_bashir}.

\begin{table}[t]
\caption{
Percentage of overlap between the different sets of trackers for the CS edges extracted from~\cite{papadopoulos2019cookie}.
We refer to the common number of trackers as \textbf{$|N|$}.
}
\label{tab:cs_pairs_tef}
\begin{center}  
\resizebox{1.0\columnwidth}{!}{  
\begin{tabular}{c|cccccc}
\toprule
\textbf{Dataset} & \textbf{$|N|$} &  \textbf{{$|E_{CS}|$}} & \textbf{$|E_{TT}|$}& \textbf{ $O_{common}$} & \textbf{ $O_{\neg CS}$   }& \textbf{ $O_{\neg TT}$} \\
\midrule
\rowcolor{Gray}
Sep17  & 226   & 3015  & 28631   & 59.70  & 49.10   & 1.70\\
Jan18   &226  & 3015  & 28631   & 59.70  & 49.10   & 1.70\\
\rowcolor{Gray}

May18   & 226   & 3024  & 27353   & 58.10  & 46.80   & 1.90\\
Jun18    & 222   & 3003  & 24943   & 55.30   & 44.70   & 2.30\\
\rowcolor{Gray}

Nov18   & 214   & 2820  & 20976   & 52.20  & 39.90   & 2.20\\
Apr19  &210   & 2929  & 18669   & 47.30   & 37.90   & 3.30\\
\bottomrule
\end{tabular}
}
\end{center}
\end{table}

\begin{table}[t]
\caption{ 
Percentage of overlap between the different sets of trackers for the CS edges extracted from~\cite{bashir2018diffusion}.
We refer to the common number of trackers as \textbf{$|N|$}.
The  $O_{\neg TT}$ value was measured 0\% across all TT graphs.
}
\label{tab:cs_pairs_bashir}
\begin{center}  
\resizebox{0.99\columnwidth}{!}{  
\begin{tabular}{c|ccccc}
\toprule
\textbf{Dataset} & \textbf{$|N|$} &  \textbf{{$|E_{CS}|$}} & \textbf{ $|E_{TT}|$}& \textbf{ $O_{common}$ } & \textbf{ $O_{\neg CS}$}\\
\midrule
\rowcolor{Gray}
Sep17  & 42    & 104   & 7321  & 80.80 & 76.40   \\
Jan18   & 41    & 102   & 7169  & 80.40& 75.90  \\
\rowcolor{Gray}

May18    & 41    & 104   & 6860  & 78.90  & 71.80  \\
Jun18    & 41    & 104   & 6389  & 73.10 & 69.20\\
\rowcolor{Gray}

Nov18    & 40    & 104   & 5896  & 69.30 & 68.50    \\
Apr19  & 39    & 104   & 5309  & 64.40 & 63.20   \\
\bottomrule
\end{tabular}
}
\end{center}
\end{table}

According to Table~\ref{tab:cs_pairs_tef}, the overlap between CS and $TT$ edges across snapshots is $47-60\%$.
In the smaller CS dataset (Table~\ref{tab:cs_pairs_bashir}), this overlap is even higher, ranging to $64-81\%$.
We remind the reader that this overlap is very high, considering all possible ways that such a graph could have been wired just by chance.
Considering that the $TT$ graphs were built artificially, this high overlap gives us an indication about the ``nature" of CS pairs, and how such data sharing flows can be found in a $TT$ graph.
This is a crucial finding: we can detect potentially collaborating pairs of trackers, without the need to deploy infrastructure to collect real users' data, or train artificial personas to collect CS activity.
Also, the overlap is reduced over time possibly because of two reasons.
First, the ground truth datasets were collected closer to our initial crawl (2017).
Second, as reported by~\cite{urban2018unwanted}, there has been a reduction in CSs after GDPR.
Therefore, the CS ecosystem is changing, and 3rd-parties that engaged in CS in earlier snapshots may not be doing so later, affecting the measured pair overlap.s
Moreover, since the first CS graph was weighted, we investigated how well $TT$ edges that overlap with CS edges cover the weights distribution.
That is, how representative $TT$ edges are of CS edges, with respect to weights.
We found that the common $TT$ edges cover well ($\sim$75\%)  of the distribution of the CS weights.
Finally, 2\% of the overlapping $TT$ edges are the 5\% most frequent edges on the CS graph, hinting again that synthetic data can capture confirmed real-world CS cases.

	\section{Discussion \& Limitations}
\label{sec:discussion}

We presented a first of its kind longitudinal study on the changes 
of the web tracking ecosystem during the last 2-3 years.
The analysis through time was performed using six crawls of the top Alexa websites, with snapshots collected a few months apart.
In the first level of analysis, we focused on network-level traffic between publishers and trackers.
We found that there are fewer trackers embedded in the websites through time, with a reduction of 9\% for the median site, and 10\% for the 90\% percentile site.
Moreover, there are fewer 3rd-party HTTP(s) requests, with  a reduction of 17\% for the median and 25\% for the 90\% percentile site.

In the second level of analysis, we constructed bipartite graphs of publishers connected with their trackers ($PT$) and studied the $PT$ graph properties 
through time.
In summary, the network structure of the tracking ecosystem and how trackers are embedded in publishers remained the same through time.
The same 3rd-parties that existed through time have been forced to cover more websites, and especially the top central trackers.
Also, top trackers in terms of publisher coverage and centrality in $PT$ graphs (e.g., \textit{google-analytics}, \textit{doubleclick}, \textit{facebook}, \textit{criteo}, \textit{appnexus}, etc.), remained top through time.
In terms of node importance, tracker sets appeared to be central in many of the examined centrality metrics.

Finally, we constructed tracker-to-tracker graphs ($TT$) for trackers who coexisted in the same websites.
We compared these models with confirmed cookie synchronization (CS) pairs, which we use as ground truth flows between tracking entities.
We found high overlap between the $TT$ edges and the CS pairs, which means that we can detect potential CS activity and data sharing flows between trackers with a practical and cheaper alternative than collecting data from real users.
Furthermore, the investigation of $TT$ graph properties uncovered ``hidden'', and not studied so far, relationships between CS and $TT$ pairs.

\noindent
\textbf{Limitations}.
The measurements presented in this paper constitute initial steps, along with related works, towards exploring the evolution of online tracking and the ad-ecosystem.
Measuring such changes and identifying the tracking patterns and behaviors of the Web is a non-trivial process, as it can involve various vantage points and types of network connections, devices, user demographics, emulated and/or real users, etc.
We believe that our study provides a lower bound on the intensity of the tracking ecosystem’s state, but more work is needed towards establishing a full and representative picture of the online tracking, and its changes through time.
Thus, our work can be extended in the following directions:

\begin{itemize}
\item \textbf{Main Page Crawls}. We performed multiple, consistent, easily replicable and lightweight- in terms of computing power and network usage- crawls on each website's main page.
Indeed different tracking may be happening in subdomains and in nested website's layers, compared to the main domain page. 
Although it is important to go beyond the main page, it is also challenging, as it requires more resources and careful planning on browsing behavior and activity exhibited by the crawlers (e.g., random vs. consistent, visit duration, etc.).
Interestingly, this aspect of the problem was covered by recent work of Urban et. al.~\cite{urban2020beyond}, where they measured the 3rd-party tracking mechanisms in subdomains of 10k popular pages.
However, they did not investigated the problem in a time-longitudinal fashion.
Thus, future work could study how subdomains' tracking ecosystem evolves through time.

\item \textbf{IP Address \& Location}. We measured the ecosystem’s tracking state using an EU-based IP address, in order to highlight the potential GDPR effect on the EU space.
Measuring biases, discrimination and other differences in tracking between locations inside and outside the EU (e.g., EU vs. USA, or EU vs. Asia) and the type of IP address used (e.g., residential, cellular, etc.) is a complex research problem in itself with several parameters at play, and it was out of the scope of this paper. 
In fact, recent works (e.g.,~\cite{zeber2020representativeness,dabrowski2019measuring}) have investigated this problem in different ways, but not in a time-longitudinal fashion.
Therefore, future work could study this tracker evolution while changing the crawling location.

\item \textbf{Cookie Synchronization Detection}. 
One could argue that our crawling method is not able to capture the full scale of the cookie synchronization ecosystem.
Indeed, measuring this kind of information exchange is challenging and requires specialized methodology such as using user synthetic personas~\cite{carrascosa2015, solomos2019ctd-raid}.
Instead, our approach is simpler and relies on automated crawls without personas, which captures much of the cookie synchronization, but not all.
Thus, it measures a lower bound of such data flows, as would be expected for real users~\cite{papadopoulos2019cookie}.
Future work could focus on utilizing more elaborated techniques on top of ours, to more accurately capture and study Cookie Synchronization flows.
\end{itemize}

\color{black}

	\section{Conclusion}
\label{sec:conclusion}
In this paper, we presented a time-longitudinal study of the online tracking ecosystem, in order to analyze its changing state over a period of two years.
We revealed patterns of the Web tracking, and effects of privacy laws such as GDPR on the tracking's intensity in over 5000 popular websites.
Apart from studying such trends, we also presented graph models easily built over our automated crawls.
These graphs capture different aspects of the tracking ecosystem, including key or central trackers with high embeddedness with respect to connectivity with other trackers, their influence on websites, as well as changes in their ranking of importance.
We also studied how information flows between trackers, due to cookie synchronization in traffic of real users, are reflected and match flows found in our automated (non-real user) crawls.
Our study points to the future opportunity for building accurate and representative graph models of the Web tracking ecosystem, without the overhead of collecting real user data.

\section*{Acknowledgements}
We thank our shepherd Ruben Cuevas and
the anonymous reviewers for their helpful feedback.
The research leading to these results has received funding from the EU's H2020 Programme under grand agreements 830927 (project CONCORDIA) and 871370 (project PIMCITY).
The paper reflects only the authors’ views and 
the Commission is not responsible fsor any use that may be made of the information it contains.

\color{black}

\bibliographystyle{IEEEtran}
\bibliography {paper.bib}

\end{document}